\documentclass[a4paper]{jpconf}
\usepackage{graphicx}

\newcommand{\Teff} {$T_{\rm eff}$}
\newcommand{\grav} {log\,{\em g}}
\newcommand{\eHe} {$\epsilon_{\rm He}$}
\newcommand{\vsini} {$v$\,sin\,$i$}
\newcommand{\macro} {$\Theta_{\rm RT}$}
\newcommand{\micro} {$\zeta_{\rm t}$}
\newcommand{\Q} {log\,$Q$}
\newcommand{\vinf} {$v_{\infty}$}
\newcommand{\kms} {km\,s$^{-1}$}

\newcommand{\ion}[2] {#1\,{\sc #2}}
\newcommand{\fastwind} {FASTWIND}

\begin{document}
\title{The IACOB project: A grid-based automatic tool for the quantitative 
spectroscopic analysis of O-stars}

\author{S. Sim\'on-D\'iaz$^{(1,2)}$, N. Castro$^{(3)}$, A. Herrero$^{(1,2)}$, 
J. Puls$^{(4)}$, M. Garcia$^{(1,2)}$, and C. Sab\'in-Sanjuli\'an$^{(1,2)}$}

\address{$^{(1)}$Instituto de Astrof\'isica de Canarias, 38200 La Laguna, Tenerife, Spain.}             
\address{$^{(2)}$Departamento de Astrof\'isica, Universidad de La Laguna, 38205 La Laguna, Tenerife, Spain.}
\address{$^{(3)}$Inst. of Astronomy \& Astrophysics, National Obs. of Athens, 15236 Athens, Greece.}
\address{$^{(4)}$Universit\"atssternwarte M\"unchen, Scheinerstr. 1, 81679 M\"unchen, Germany.}

\ead{ssimon@iac.es}

\begin{abstract}
We present the IACOB grid-based automatic tool for the quantitative spectroscopic 
analysis of O-stars. The tool consists of an extensive grid of FASTWIND models, and a 
variety of programs implemented in IDL to handle the observations, perform the automatic analysis, 
and visualize the results. The tool provides a fast and objective 
way to determine the stellar parameters and the associated uncertainties of large 
samples of O-type stars within a reasonable computational time.
\end{abstract}

\section{Introduction}\label{intro}
The era of large spectroscopic surveys of massive stars has already begun, providing us with a huge amount of 
high-quality spectra of Galactic and extragalactic O-type stars. The {\em IACOB spectroscopic 
survey of Northern Galactic OB-stars} is one of them. This long-term observational project is aimed 
at building a multi-epoch, homogeneous spectroscopic database of high-resolution, high 
signal-to-noise ratio spectra of Galactic bright OB-stars\footnote{Some notes on the present status of 
{\em IACOB spectroscopic database}, the IACOB project, and the various working packages can be found 
in [1] and [2].}. Associated with this 
spectroscopic dataset are several working packages aimed at its scientific exploitation. 
Within the framework of the Working Package {\em WP3: Quantitative Spectroscopic Analysis}, we have 
developed a powerful tool for the automatic analysis of optical spectra of O-type stars.
The tool provides a fast and objective way to determine the stellar parameters and the associated 
uncertainties of large samples of O-stars within a reasonable computational time.
Initially developed to be used for the analysis of spectra of O-type stars from the {\em IACOB spectroscopic database},
the tool is now also being applied in the context of the {\em VLT-FLAMES Tarantula Survey} (VFTS) project
[3], and other studies of stars of this type.
\section{The IACOB grid-based tool}\label{tool}
Apart from the already mentioned characteristics (automatic, objective, and fast), we also aimed at a
tool that is portable, versatile, adaptable, extensible, and easy to use. As shown throughout the following text,
this philosophy has guided the whole development of the tool.\\
\newline
One of the key-stones of any automatic quantitative spectroscopic analysis is the computation of
large samples of synthetic spectra (using a stellar atmosphere code), to be compared with the
observed spectrum. In contrast to other possible alternatives (the {\em genetic algorithm} --\,GA\,-- employed
by [4], or the {\em principal component analysis} --\,PCA\,-- approach followed by
[5]), we decided to base our automatic tool on an extensive, precomputed grid of stellar 
atmosphere models and a line-profile fitting technique (i.e., a grid-based approach\footnote{
A similar approach has been recently applied by [6] for the analysis of a sample of 12 low resolution
spectra of B supergiants in NGC55.})
This option was 
possible thanks to the fast performance of the \fastwind\ code\footnote{Among the various available 
modern stellar atmosphere codes incorporating wind and line-blanketing effects, \fastwind\ [7]
is the fastest.}, and the availability of a 
cluster of $\sim$\,300 computers at the Instituto 
de Astrof\'isica de Canarias (IAC) connected through the CONDOR\footnote{http://www.cs.wisc.edu/condor/} 
workload management system.\\
\newline
The grid is complemented by a variety of programs implemented in IDL, to handle the observations, 
perform the automatic analysis, and 
visualize the results. The IDL-package has been built
in a modular way, allowing the user, e.g., to easily modify how the mean values 
and uncertainties of the considered parameters are computed, or which evolutionary tracks will be used 
to estimate the evolutionary masses.
\subsection{The IACOB FASTWIND grid}\label{grid}
In the following we outline the main characteristics of the FASTWIND grid used for the analysis of Galactic O-type stars 
that is presently incorporated within the IACOB grid-based tool.\\
\newline
{\em Input stellar parameters considered in the FASTWIND models:}
\begin{itemize}
 \item {\bf Effective temperature (\Teff) and gravity (\grav):} Table \ref{grid_table} indicates 
the ranges of \Teff\ and \grav\ considered in the grid. Basically, 
the grid points were selected to cover the region of the log\,\Teff\,--\,\grav\, diagram 
where the O-type stars are located. 
 \item {\bf Helium abundance (\eHe) and microturbulence (\micro)}: 
The grid includes six values of
helium abundance --- \eHe\,=\,N$_ {\rm He}$/(N$_ {\rm H}$+N$_ {\rm He}$) ---, indicated in Table \ref{grid_table}.
For all models, a microturbulence \micro$_{\rm , mod}$\,= \,15\,\kms\ was adopted in the computation of the 
atmospheric structure, 
and four values of the microturbulence were considered in the formal solution.   
 \item {\bf Radius ($R$):} Computing a \fastwind\ model requires an input value for the radius.
This radius has to be close to the actual one (which will be derived in the final step of the 
analysis, and hence is not know from the beginning on). We assumed a radius for each
(\Teff, \grav) pair following the calibration by [8]. The grid is hence divided in 20 regions
in which a different radius is considered. For example, for the case of \grav\,=\,4.0 and 3.5 dex, the radii range from 
7 to 12 $R_{\odot}$ and from 19 to 22 $R_{\odot}$, respectively.
 \item {\bf Wind parameters ($\dot{M}$, \vinf, $\beta$):} As conventional for grid computations of stellar
atmosphere models for the optical analysis of O-stars, the mass loss rate 
($\dot{M}$) and terminal velocity (\vinf) of the wind have been compressed, together with the radius, into 
the wind strength parameter (or optical depth invariant), $Q = \dot{M}/(v_{\infty} R)^{1.5}$ (see [7]). 
Ten log\,$Q$-planes 
were considered for the grid. For each FASTWIND model, $\dot{M}$ and \vinf\ need to be specified. First,
a terminal velocity \vinf\,=\,2.65\,$v_{\rm esc}$ was adopted, following [9]; 
then a mass loss rate was computed for the
given \Q, \vinf, and $R$. 
Finally, the exponent of the velocity law, $\beta$,
was assumed as a free parameter ranging from 0.8 to 1.8.
 \item {\bf Metallicity}: A solar metallicity (following [10]) was assumed for the 
whole grid\footnote{We also calculated similar grids for other metalicities 
($Z$\,=\,0.5, 0.4, and 0.2 $Z_{\odot}$).}.\\
\end{itemize}
{\em Synthetic lines:} 
The following (optical) lines where synthesized in the formal solution: H$_{\alpha-\epsilon}$, 
\ion{He}{i}\,$\lambda\lambda$4026, 4120, 4143, 4387, 4471, 4713, 4922, 5015, 5048, 5875, 6678, and 
\ion{He}{ii}\,$\lambda\lambda$4026, 4200, 4541, 4686, 5411, 6406, 6527, 6683.
\begin{figure}[t!]
\begin{minipage}[m]{22.0pc}
\includegraphics[height=21.5pc, angle=90]{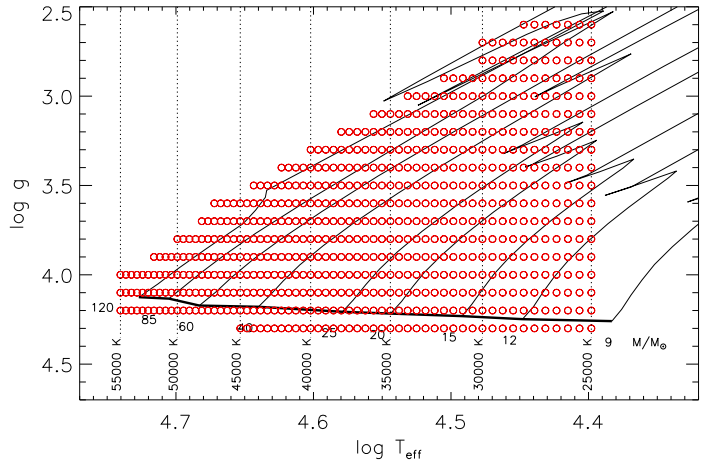}%
\end{minipage}
\hspace{1.0pc}
\begin{minipage}[m]{14.5pc}
\caption{\label{grid_figure}Location of (\Teff, \grav)\,-\,pairs considered in the
FASTWIND grid incorporated into the present version of the IACOB grid-based tool. 
Evolutionary tracks from [11].}
\end{minipage}
\end{figure}
\begin{center}
\begin{table}[t!]
\centering
\caption{\label{grid_table}Range of values considered for the six free parameters in the FASTWIND grid.} 
\begin{tabular}{r l}
\br
\grav\,: & [2.6\,--\,4.3]\,dex (with step size 0.1 dex)\\
\Teff\,: & $\ge$\,25000\,K (step size 500\,K), upper limit defined by the 120 M$_{\odot}$ track (see Fig. \ref{grid_figure})\\
\micro\,: & Model: 15 \kms, Formal solution: 5, 10, 15, 20 \kms \\
\eHe\,: & 0.06, 0.09, 0.13, 0.17, 0.20, 0.23 \\
\Q\,: & -15.0, -14.0, -13.5, -13.0, -12.7, -12.5, -12.3, -12.1, -11.9, -11.7 \\
$\beta$\,: & 0.8, 1.0, 1.2, 1.5, 1.8 \\
\br
\end{tabular}
\end{table}
\end{center}
In order to optimize the size and the read-out time of the grid, only part of the output 
from the \fastwind\ models is kept and stored in IDL xdr-files. This includes the input 
parameters, the H/He line profiles and equivalent widths, the information 
about the stellar atmosphere structure and the emergent flux distribution, 
and the synthetic photometry resulting from the computed emergent flux distribution. Using 
these xdr-files, the size of the grid can be reduced to a 10\,\% of the original one. 
IDL can restore each xdr-file and compute the corresponding $X^2_T$ quantity (see Sect. \ref{idl}) 
in 0.02\,--\,0.1 s per model,
i.e., the tool can pass through 80,000 models in 30 min\,--\,1 hour. Following the main philosophy of the
IACOB grid-tool, only
the reduced grid is used by the tool,
whilst the original grid is safely kept on hard-disk at the IAC. This allows for an easy
transfer of the grid to an external disk, hence satisfying our constraint for portability.
\\ \newline
The \fastwind\ grid presently incorporated into the IACOB grid-based tool consists of
$\sim$\,192,000 models ($\sim$\,32,000 models per He-plane). The reduced size is $\sim$\,34 Gb. 
This predefined grid can be easily updated and/or extended if necessary 
using appropriate scripts implemented in IDL, CONDOR, and LINUX. For example, a new He-plane
can be computed and prepared for use in $\sim$\,3 days.
\subsection{The IACOB grid-based tool IDL package}\label{idl}
\begin{figure}[t!]
\center
\includegraphics[width=14.cm]{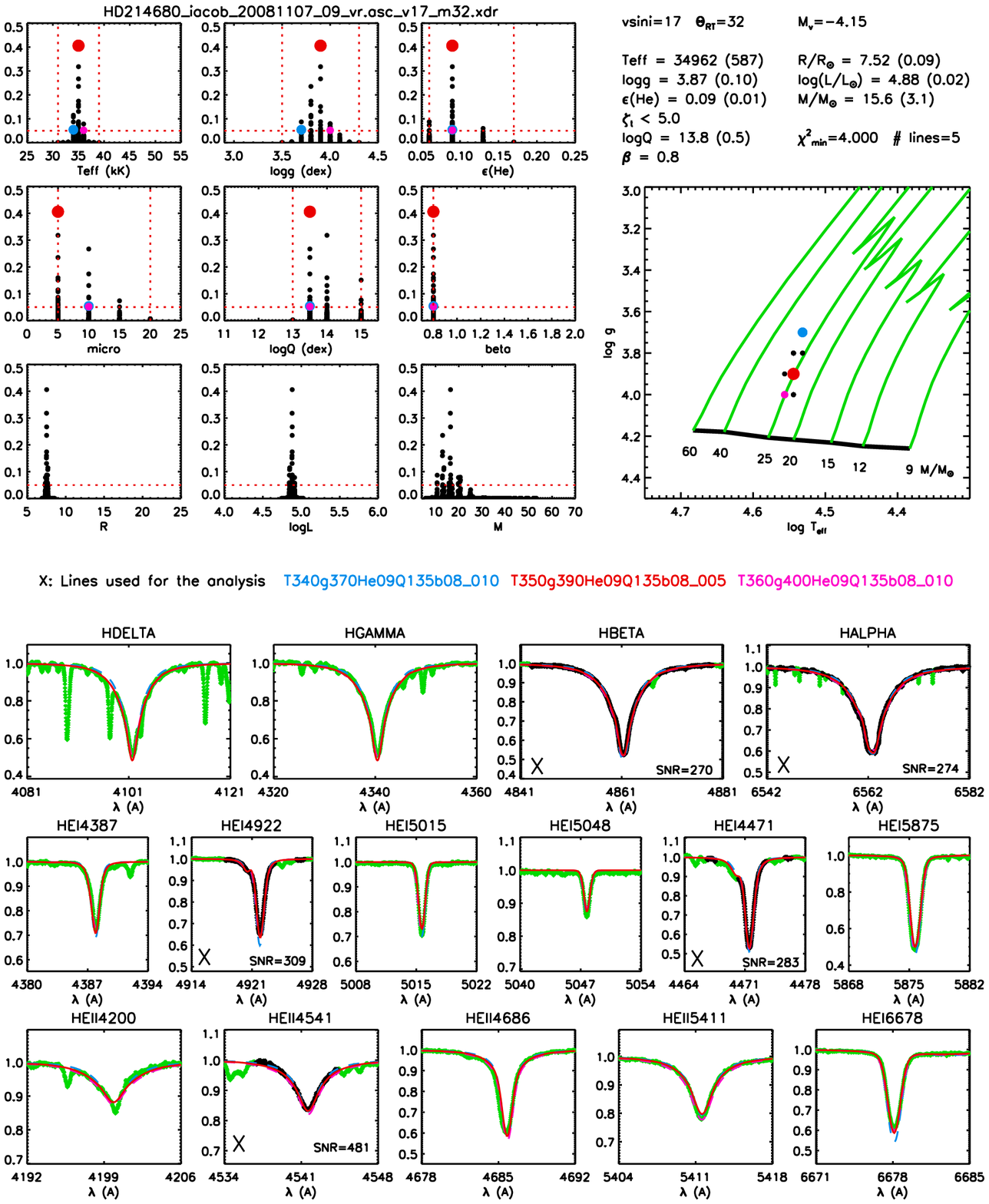}
\caption{Example of output provided by the IACOB grid-based tool, for
the quantitative spectroscopic analysis of the Galactic O9\,V star HD\,214680. 
See text for the explanation of all the information contained in the figure.
Observed spectrum from the {\em IACOB spectroscopic database} ($R$\,=\,46,000).
\vsini\ and \macro\ were previously determined using a combined 
Fourier transform + goodness of fit technique (see e.g. [14]). 
In this example, only 5 H/He lines were used for the actual analysis. Note that
also all other synthetic lines fit the observations perfectly at the
derived parameters. The analysis was performed in $\sim$\,2 min. 
See [2] for details on the distance, apparent visual magnitude 
and visual extinction used to determine the absolute visual magnitude ($M_{\rm v}$).
\label{fig2}}
\end{figure}

The guideline of the automatic analysis is based on standard techniques applied within the 
quantitative spectroscopic analysis of O-stars using optical H/He lines that have been 
described elsewhere (e.g. [12], [13]). 
The whole spectroscopic analysis is performed by means of a variety of IDL programs following
the steps indicated below. In brief, once the observed spectrum is processed, the tool 
obtains the quantity $X_T^2$ (i.e. an estimation of the goodness of fit) for each model within 
a subgrid of models selected from the
global grid, and determines the stellar parameters and their associated uncertainties by interpreting the
obtained $X_T^2$ distributions.\\
\newline
{\em \underline{Step 1}: The input} 
\\ \newline
In this first step, the user has to provide the observed spectrum, to indicate
the corresponding resolving power ($R$), and to give some pre-determined information about the star
-- the projected rotational velocity (\vsini), the size of the 
extra line-broadening (\macro), and the absolute visual magnitude ($M_{\rm V}$).\\
Concerning the grid, the user must select the appropriate metallicity,
and indicate the range of values for the various free parameters (defining the
subgrid of models to be considered within the analysis). This latter option allows 
the tool to be faster (using optimized ranges for the various free parameters), or 
the user to perform a preliminary quick analysis by fixing some of the parameters. For example, 
one can obtain a quick estimate on \Teff, \grav, and \Q\ in less than 5 min, by fixing the other three 
parameters (\micro\, \eHe, and/or $\beta$) .\\
Finally, the H/He lines which shall be considered in the analysis and the corresponding weights 
need to be specified.
\\ \newline
{\em \underline{Step 2}: Processing of the observed spectrum} 
\\ \newline
In many cases, the observed spectra need to be processed before launching the automatic
analysis, because of, e.g., nebular contamination affecting the cores of the H and \ion{He}{i}
lines, the need to improve the normalization of the continuum adjacent to the line, and
the presence of cosmic rays.\\
The possibility to correct the observed spectrum for these effects has been incorporated
into corresponding IDL procedures. The implemented options include (for each of the considered lines): 
local renormalization, selection of the wavelength range of the line, clipping/restoring of certain 
parts of the line, and computation of the signal-to-noise ratio (S/N) in the 
adjacent continuum.\\
Once finalized, the processed spectrum is stored. This spectrum and all associated information
is used in {\em Step 3}, and also each time the analysis of the same star/spectrum is (re-)launched. 
This way the user can easily reconsider his/her decisions after a first analysis.\\
In this step, the user can select a model from the grid to help him/her with the processing
of the observed spectrum. One can, for example, make a quick preliminary analysis (see above)
fixing some parameters and including only few lines, and then use the resulting best model
to check the processed lines and 
find constraints for
the processing of the others. 
\\ \newline
{\em \underline{Step 3}: Estimation of the goodness of fit for each considered model} 
\\ \newline
This step, together with {\em Step 4}, are the most important ones, constituting the core of
the program. In the present version of the IACOB grid-based tool we have considered
a procedure as described below; however, we are aware that this procedure can be subject to discussion/improvements.
Having this in mind, and following the philosophy of the tool (regarding versatility
and adaptability), 
the corresponding IDL modules have been implemented with the possibility to be easily
modified. The fast performance of the IACOB grid-based tool makes it very powerful
to investigate various alternative strategies.\\ 
\newline
In our present version, the tool computes, for every model in the subgrid, the quantity $X^2_L$ 
for each considered line 
\begin{equation}
 X^2_L\,=\,\frac{1}{N_{\lambda}}\sum_{\lambda=1}^{N_{\lambda}}\,\frac{(F_{m,\lambda}\,-\,F_{o,\lambda})^2}{\sigma_L^2} 
\end{equation}
where $F_{m,\lambda}$ and $F_{o,\lambda}$ are the normalized fluxes corresponding to the synthetic
and observed spectrum, respectively; $\sigma_L$\,=\,(S/N)$^{-1}$ accounts for the S/N of the line; 
and $N_{\lambda}$ is the number of frequency points in the line. Under ideal conditions (e.g., for a
perfect model, but see below), $X^2_L$ corresponds to a reduced $\chi^2$.\\
\newline
In a second step, the $X^2_L$ values for each model are corrected for possible deficiencies in the
synthetic lines (due, for example, to deficiencies in the model, an incorrect characterization of the
noise of the line, bad placement of the continuum, or bad characterization of the line-broadening).
To this aim we compute, for each line, the standard deviation of the residuals 
$\sigma_c$\,=\,$\sigma$($F_{m,\lambda}$\,--\,$F_{o,\lambda}$) from that model that results in 
the minimum $X^2_L$ for the given line. Then, the following correction is applied:
\begin{equation}
 X^2_{L,c}\,=\,X^2_L\,\frac{\sigma^2_L}{\sigma^2_c}
\end{equation}
Using these individual $X^2_{L,c}$ values and the weights assumed for each of the considered 
lines ($w_L$, e.g., [4]), a global $X^2_T$ is obtained:
\begin{equation}
 X^2_T\,=\,\sum_{L=1}^{N_L}\,w_L\,X^2_{L,c} 
\end{equation}
where $N_L$ is the number of lines. For a large number of frequency points per line, $X^2_T$ 
should be normally distributed.\\
\newline
Thus, {\em Step 3} provides the values of the quantity $X_T^2$ associated with each of the models
included in the considered subgrid. As indicated in Section \ref{grid}, this step can last from a few seconds
to less than about 1~hour, depending on the number of models in the subgrid. An example of $X_T^2$\,-\,distributions 
(actually, $p_{X_T^2}$\,=\,e$^{-0.5 X_T^2}$) with respect to the various stellar parameters is presented in Figure \ref{fig2}.
\\ \newline
{\em \underline{Step 4}: Computation of mean values and uncertainties} 
\\ \newline
The previous step provides the $X^2_T$\,--\,distributions for each of the six parameters 
derived through the spectroscopic analysis (\Teff, \grav, \eHe, \micro, \Q, and $\beta$). If
the absolute visual magnitude is provided, the tool automatically determines $R$, 
log\,$L$, and $M_{\rm sp}$ (spectroscopic mass) for each model, and hence the 
corresponding $X^2_T$\,--\,distributions for these parameters are available as well.
If also the terminal velocity is provided, a similar computation is performed for the
mass-loss rate (using $Q$, $R$, and \vinf). Finally, one 
of the IDL modules computes the evolutionary masses ($M_{\rm ev}$) of the
models from an interpolation in the (log\,\Teff, \grav)-plane using the tracks
provided by a stellar evolution code. Since the $X^2_T$\,--\,distributions for
\Teff\ and \grav\ are computed and stored in {\em Step 3}, the computation of
the corresponding distributions for $R$, log\,$L$, $M_{\rm sp}$, $M_{\rm ev}$ 
can be easily repeated in a few seconds, in case a different $M_{\rm v}$ and/or
evolutionary tracks need to be considered.\\
These distributions are then used to compute mean values and uncertainties for 
each parameter (taking into account that models above a given threshold in $X^2_T$
can be discarded).\\
\newline
The tool also allows to easily investigate possible degeneracies (for example, for stars
with weak winds the analysis of the optical H/He lines only provides an upper limit
for the $Q$ parameter, and does not allow to constrain $\beta$), and the contribution 
of the various parameters to the final uncertainty (by fixing one of the free parameters 
to a certain value and recomputing the statistics for the other parameters).
\\ \newline
{\em \underline{Step 5}: Visualization of the results} 
\\ \newline
The last step is the creation of a summary plot for better visualization of the results (see 
Figure \ref{fig2} as an example). The present version of the tool includes:
\begin{itemize}
 \item  $p_{X_T^2}$\,--\,distributions for the various parameters involved in the analysis (upper-left panels).
In those panels, vertical lines indicate the limiting values adopted for the six free parameters 
(i.e., the sub-grid of FASTWIND models for which a $X_T^2$ is computed), and horizontal lines 
indicate the $p_{X_T^2}$ value corresponding to $X^2_T$(threshold). The user can select three of the models 
that will be indicated as red, blue, and pink dots (the same colors are also used in other parts
of the figure). 
\item a summary of the input parameters (\vsini, $\Theta_{\rm RT}$, and $M_{\rm v}$) and the 
result from the statistics for each parameter resulting from the analysis (upper-right part of
the figure).
\item a log\,\Teff\,---\grav\ diagram including evolutionary tracks (e.g., from [11], 
without rotation, in the figure) and the position of all models with 
$X_T^2$\,$\le$\,$X^2_T$(threshold).
\item a set of panels (lower part of the figure) were the observed and synthetic H/He lines 
indicated by the user are compared. The spectral regions that are used to compute the $X_T^2$ 
for each line are indicated in black, while the clipped (or not used) regions are 
presented in green. Note that the plotted lines are not limited to those used in the 
analysis process (the latter are marked with an 'X').
\end{itemize}

\subsection{Problems that can be investigated with the IACOB grid-based tool}
The fast performance and versatility of the IACOB grid-based tool not only allows to
analyze large samples of O-type stars in a reasonable time,
but also to easily investigate the effects of
\begin{itemize}
 \item {\em the assumed line-broadening}: The common strategy in previous spectroscopic analyses of
O-stars was to assume pure rotational line-broadening (in addition to
natural, thermal, Stark/collisional and microturbulent broadening included in the synthetic spectra). 
However, recent studies have shown that an important extra-broadening contribution (commonly
quoted as {\em macroturbulent broadening}) affects the shape of the line-profiles of
this type of stars (see e.g. [15], and references therein). How much are the 
derived parameters affected when this extra-broadening is
neglected? Can both broadening contributions be represented by one 'fake' rotational profile ($v_f$\,sin$i$),
without disturbing the resulting parameters? What is the effect of the uncertainty in the 
derived/assumed broadening? Some results from a preliminary investigation of these questions
can be found in [16].
 \item {\em the placement of the continuum}: The normalization of the spectra of O-type stars is
sometimes complicated, especially in the case of echelle spectra. It is commonly argued that
the derived gravity can be severely affected by the assumed normalization, but, to which extent? 
To investigate this effect one could use the IACOB grid-based tool, applying small 
modifications to the continuum placement. 
An example of this type of analysis for the case of low resolution spectra of B-supergiants 
can be found in [6].
 \item {\em neglecting/including a variety of different commonly used diagnostic lines}: Before computing the
global $X^2_T$ (see equation 3), the resulting $X^2_L$ for each line are stored. This way
the user can easily recompute $X^2_T$, discarding some of the initially considered lines. This
option allows, for example, to investigate the change in the wind-strength parameter when
both H$\alpha$ and \ion{He}{ii}\,4686 lines or only one of them are included in the analysis.  
 \item {\em clipping part of the lines}: Most of the O-stars are associated to H\,{\sc ii} regions. In some cases, the stellar
spectra can be heavily contaminated by nebular emission lines (mainly the cores of the
hydrogen lines, but also the \ion{He}{i} lines). This contamination must be identified and eliminated
from the stellar spectrum to obtain meaningful results from the spectroscopic analysis. When the resolution
of the spectra is high, the clipped region is small in comparison with the total line width; however,
even for a moderate resolution an important region of the line needs to be eliminated. What is the effect of
such a strong clipping in the H$\alpha$ line on the determination of the mass-loss rate?
How large is the effect of clipping the core of the \ion{He}{i} lines in the \Teff\ determination?
 \item {\em the way the statistics is derived from the $X^2_T$\,--\,distributions}: As commented in Section \ref{idl},
this is a critical point in the determination of final best values and uncertainties.
\end{itemize}
These are only some examples of things can be investigated with the IACOB grid-based tool. Many other
tests can be easily performed. In addition, we expect the tool to be of great benefit for the analysis 
of the O-star samples included in, e.g., the ESO-Gaia, VFTS, OWN, IACOB and other similar large surveys.
 
\ack
Financial support by the Spanish Ministerio de Ciencia e
Innovaci\'on under projects 
AYA2008-06166-C03-01,	
AYA2010-21697-C05-04,
and by the Gobierno de Canarias under project 
PID2010119.
This work has also been partially funded by the Spanish MICINN under the 
Consolider-Ingenio 2010 Program grant CSD2006-00070: First Science with the GTC
({\tt http://www.iac.es/consolider-ingenio-gtc}).%

\end{document}